\documentclass[twocolumn,showpacs,prd,amsmath,amssymb,nofootinbib,superscriptaddress]{revtex4-2}
\usepackage[utf8]{inputenc}
\usepackage{color}
\usepackage{float}
\usepackage{amsmath}
\usepackage{amssymb}
\usepackage{graphicx}
\PassOptionsToPackage{version=3}{mhchem}
\usepackage{mhchem}
\usepackage[unicode=true,
 bookmarks=false,
 breaklinks=true,pdfborder={0 0 1},backref=section,colorlinks=true]
 {hyperref}
\hypersetup{
 pdfstartview=Fit,linkcolor=blue,citecolor=blue,urlcolor=blue}

\makeatletter
\pdfoutput=1
 
\usepackage{cases}
\usepackage{amsfonts}
\usepackage{dcolumn}
\usepackage{bm}
\usepackage{bbm}
\usepackage{xcolor}
\usepackage{array}
\usepackage{subfigure}
\usepackage{wasysym}
\usepackage{dsfont}
\usepackage{booktabs}

\newcommand{\be}{\begin{equation}}
\newcommand{\ee}{\end{equation}}
\newcommand{\ba}{\begin{eqnarray}}
\newcommand{\ea}{\end{eqnarray}}

\newcommand{\gsim}{\mathrel{\hbox{\rlap{\lower.55ex \hbox {$\sim$}}
                   \kern-.3em \raise.4ex \hbox{$>$}}}}
\newcommand{\lsim}{\mathrel{\hbox{\rlap{\lower.55ex \hbox {$\sim$}}
                   \kern-.3em \raise.4ex \hbox{$<$}}}}

\makeatother

\begin{document}
\title{New models and Big Bang Nucleosynthesis constraints in $f(Q)$ gravity}

\author{Fotios K. Anagnostopoulos }
\email{fotis-anagnostopoulos@hotmail.com}
\affiliation{Department of Informatics and Telecommunications, University of Peloponnese, Karaiskaki 70, 22100, Tripoli, Greece}
\author{Viktor Gakis}
\email{vgakis@central.ntua.gr}

\affiliation{Department of Physics, National Technical University of Athens, Zografou
Campus GR 157 73, Athens, Greece}
\author{Emmanuel N. Saridakis}
\email{msaridak@phys.uoa.gr}

\affiliation{National Observatory of Athens, Lofos Nymfon, 11852 Athens, Greece}
\affiliation{Department of Astronomy, School of Physical Sciences, ~~\\
 University of Science and Technology of China, Hefei, Anhui 230026,
China}
\affiliation{CAS Key Laboratory for Research in Galaxies and Cosmology, ~~\\
 University of Science and Technology of China, Hefei, Anhui 230026,
China}
\author{Spyros Basilakos}
\email{svasil@academyofathens.gr}
\affiliation{National Observatory of Athens, Lofos Nymfon, 11852 Athens, Greece}
\affiliation{Academy of Athens, Research Center for Astronomy and Applied Mathematics,
Soranou Efesiou 4, 11527, Athens, Greece}
\affiliation{School of Sciences, European University Cyprus, Diogenes Street, Engomi, 1516 Nicosia, Cyprus}


\begin{abstract}
   The $f(Q)$ theories of modified gravity arise from the consideration of 
non-metricity as the basic geometric  quantity, and have been proven to be very 
efficient in describing the late-time Universe. We use the Big Bang 
Nucleosynthesis (BBN) formalism and observations in order to 
extract constraints on  various classes of $f(Q)$ models. In particular, we  
 calculate the deviations that $f(Q)$ terms bring on the freeze-out 
temperature $T_f$ in  comparison  to that of the standard 
$\Lambda$CDM evolution, and then we impose the observational  bound on $ 
\left|\frac{\delta {T}_f}{{T}_f}\right|$   to extract constraints on the 
involved parameters of the considered  models. 
Concerning the  polynomial model, we show that the exponent parameter 
 should be negative, while for the power-exponential model and the new    
hyperbolic tangent - power model we find
that they pass the BBN constraints trivially.   Finally, we 
examine   two    DGP-like $f(Q)$
models, and we extract the bounds on their model parameters.
Since many gravitational modifications, although able to 
describe the late-time evolution of the Universe, produce too-much modification 
at early times and thus  fall to pass the BBN 
confrontation, the fact that  $f(Q)$ gravity can safely pass the BBN 
constraints is an important advantage of this modified gravity class.

\end{abstract}

\maketitle

\section{Introduction}

The accelerated expansion of the late Universe is commonly attributed
to the existence of an extra cosmic constituent known collectively as ``Dark 
Energy''. In the context of the concordance   $\Lambda$CDM scenario, dark
energy is just the standard cosmological constant and this scenario
is very successful in interpreting observations such as Cosmic Microwave 
Background (CMB) \cite{Aghanim:2018eyx},
Supernovae Ia (SNIa) \cite{scolnic2018complete}, Large Scale Structure 
(LSS)
\cite{Gil-Marin:2015sqa}, etc. 

However, recently various observational
issues have gain attention, for example the Hubble constant tension, that 
is
the discrepancy between the value of the Hubble constant from the
CMB and the corresponding one from Cepheids  up to ~5$\sigma$ (see
\cite{di2020cosmology} for a thorough and recent review). Another
example is the difference of the root-mean-square amplitude of matter 
over-density perturbation $\sigma_{8}$, between the value extracted from the CMB
(imposing $\Lambda$CDM) and the corresponding value extracted by
fitting on the Large Scale power spectrum  \cite{di2021cosmology}.
Furthermore, the value of the cosmological constant in the late Universe
seems to be inconsistent with the corresponding value at the early
Universe  \cite{weinberg1989cosmological}. 

On the other hand, the current
  gravitational theory, namely General Relativity (GR), cannot be
 re-normalized and thus it is difficult to be consistent with a quantum 
description \cite{Addazi:2021xuf}. Hence, a large portion of the literature 
follows the direction of modified gravity, which seems
promising to solve the latter problem  \cite{stelle1977renormalization}, and in 
addition it is very efficient in describing the 
two phases of accelerated expansion of the universe 
\cite{CANTATA:2021ktz,Capozziello:2011et}. Finally, note that modified gravity 
can be efficient in alleviating  the two aforementioned tensions of $\Lambda$CDM
cosmology, namely the $H_{0}$ and the $\sigma_{8}$ ones \cite{Abdalla:2022yfr}.

Along these lines and due to the fact that the Einstein-Hilbert
Lagrangian is just one of many possible Lagrangians that could lead
to the same field equations, the community has developed many frameworks
of modified gravity. A class of gravitational modification
arises from the extension of the Einstein-Hilbert Lagrangian with
extra terms, however maintaining its geometrical structure, i.e. its
Riemannian formulation, resulting for instance in $f(R)$ gravity \cite{DeFelice:2010aj},
in Gauss-Bonnet and $f(G)$ gravity \cite{Antoniadis:1993jc,Nojiri:2005jg},
in Weyl gravity \cite{Mannheim:1988dj}, etc. A more radical approach
to follow is to modify the building blocks of General Relativity,
that is the underlying geometrical structure. An instance
of this kind of modification arises from the
torsional formulation of gravity, namely the modified teleparallel
theories, such as the $f(T)$ gravity \cite{Bengochea:2008gz,Cai:2015emx},
the $f(T,T_{G})$ gravity \cite{Kofinas:2014owa}, the scalar-torsion
theories \cite{Geng:2011aj,Hohmann:2018rwf} etc. In contrast to the standard GR
and the corresponding $f(R)$ extensions, which are based on curvature,
in modified teleparallel formulation 
gravity is manifested via the torsion tensor, with curvature being 
zero.

There is yet another possibility, which is formulated using 
the non-metricity $Q$  of the connection. In this setup, only the
non-metricity is non-zero while curvature and torsion both vanish. Hence, 
non-metricity can be used in order to describe gravity through geometry, 
leading to   the so-called Symmetric Teleparallel Equivalent to General 
Relativity (STEGR)   \cite{BeltranJimenez:2017tkd}.
  It is interesting to note that although Teleparallel 
Equivalent
of General Relativity (TEGR) as well as STEGR, are completely equivalent to 
General Relativity at the level of equations, when we modify them
the resulting theories $f(R)$, $f(T)$ and $f(Q)$ are   different 
theories. These types of 
modifications have in general more degrees of freedom than the two of standard 
GR, unless specific constraints are imposed in the functional forms.  Thus, 
during the last years $f(Q)$
gravity, its application to cosmology and generalizations has attracted
the interest of the literature \cite{Jimenez:2019ovq,Xu:2019sbp,Jimenez:2019ghw,Percacci:2019hxn,Delhom:2020vpe,Delhom:2020hkb,Barros:2020bgg,BeltranJimenez:2020sih,Jimenez-Cano:2020chm,DAmbrosio:2020nev,Rubiera-Garcia:2020gcl,Barros:2020bgg,BeltranJimenez:2020sih,Jimenez-Cano:2020chm,Xu:2020yeg,Ayuso:2020dcu,Cabral:2020fax,Flathmann:2020zyj,Frusciante:2021sio,Yang:2021fjy,Fu:2021rgu,Khyllep:2021pcu,Anagnostopoulos:2021ydo,Albuquerque:2022eac}. 

A very strong constraint on modified gravity models in the early Universe
comes from the era of Big Bang Nucleosynthesis (BBN) 
\cite{Bernstein:1988ad,Kolb:1990vq,Olive:1999ij,Cyburt:2015mya,Barrow:2020kug, 
Asimakis:2021yct}.
 In the present work we are interested in  imposing
BBN constraints on $f(Q)$ gravity, within an  analytical  approach,
following \cite{Torres:1997sn,Lambiase:2005kb,Lambiase:2012fv,Lambiase:2011zz,Capozziello:2017bxm}. Specifically, we use observational
data on primordial abundances of $\ce{^{4}He}$ to constrain all known
$f(Q)$ models along some that are proposed for the first time inspired by the 
successful
$f(T)$ models. The plan of the manuscript is the following. In Section 
\ref{Sec1} we provide a formal
description of Symmetric Teleparallel gravity and its $f(Q)$ extension, while 
in Section \ref{sec:Cosmological-application-of} we present the cosmological 
solutions
of the theory alongside the  specific  $f(Q)$ models that are going to be 
studied. In Section \ref{Sec3} we briefly review the formalism of BBN 
constraints, and then in Section \ref{Sec4} we apply it in the framework of 
$f(Q)$ gravity.   Finally, in Section \ref{Conclusions} we derive our 
conclusions and we point out further directions.

\section{Symmetric Teleparallel Gravity}
\label{Sec1}

General Relativity is built upon Lorentzian Geometry which is specified by 
choosing a connection that is symmetric and metric compatible. The latter is the
Levi-Civita connection and due to its properties it only produces non-zero 
curvature, whilst torsion and non-metricity are zero \cite{Nakahara:2003nw}. 
Nevertheless, a different kind of connection can be utilized when using 
geometrodynamics as the mathematical framework for gravity. As a matter of fact 
the most general connection is called metric-affine it is described by the 
formula
\begin{align}
\Gamma^{\alpha}{}_{\mu\nu} & =\mathring{\Gamma}^{\alpha}{}_{\mu\nu}+K^{\alpha}{}_{\mu\nu}+L^{\alpha}{}_{\mu\nu}\,,\label{eq:connection_expansion}
\end{align}
where $\mathring{\Gamma}^{\alpha}{}_{\mu\nu}$ represents the Christoffel
symbols of the Levi-Civita connection, $K^{\alpha}{}_{\mu\nu}$ is
the contorsion tensor and $L^{\alpha}{}_{\mu\nu}$
is the \textit{disformation} tensor related to the non-metricity 
\cite{Nester:1998mp}
\begin{align}
Q_{\alpha\mu\nu} & :=\nabla_{\alpha}g_{\mu\nu}\,.\label{eq:Qtensor}
\end{align}
The disformation tensor can then be expanded as \cite{BeltranJimenez:2017tkd}
\begin{align}
L^{\alpha}{}_{\mu\nu} & 
:=\frac{1}{2}g^{\alpha\beta}\left(-Q_{\mu\beta\nu}-Q_{\nu\beta\mu}+Q_{
\beta\mu\nu}\right)\,,\label{eq:Ltensor}
\end{align}
and encapsulates all the information coming from the non-metricity tensor of 
the general
affine connection. In summary, these three types of geometric deformation 
$\mathring{\Gamma}^{\alpha}{}_{\mu\nu}$, $K^{\alpha}{}_{\mu\nu}$ and 
$L^{\alpha}{}_{\mu\nu}$
form a kind of ``trinity of gravity'', which encompasses all components of the 
general
connection $\Gamma^{\alpha}{}_{\mu\nu}$ \cite{BeltranJimenez:2019tjy}. This 
practically implies that it is possible to express the geometry of a 
gravitational theory in either curvature, non-metricity or torsion. A connection 
which admits only non-metricity, whilst curvature and torsion are zero, is 
called Symmetric Teleparallel Gravity (STG) connection, while
connections which have only zero curvature are called teleparallel connections 
\cite{ortin2007gravity,BeltranJimenez:2018vdo}.

Similarly to  TEGR, which is based on a connection 
which has only torsion 
\cite{aldrovandi2012teleparallel},   the STG connection  has only 
non-metricity  and hence STEGR is equivalent to GR  at the level of field 
equations, since its action differs by a boundary term from the 
Einstein-Hilbert. In particular, defining the non-metricity scalar  as 
\cite{BeltranJimenez:2017tkd} 
\begin{equation}
Q:=-\frac{1}{4}Q_{\alpha\mu\nu}Q^{\alpha\mu\nu}+\frac{1}{2}Q_{\alpha\mu\nu}Q^{
\mu\alpha\nu}+\frac{1}{4}Q_{\alpha}Q^{\alpha}-\frac{1}{2}\bar{Q}_{\alpha}Q^{
\alpha}\,,\label{eq: Q_GR}
\end{equation}
with
\begin{align}
Q_{\mu} & 
:=Q_{\mu}{}^{\alpha}{}_{\alpha}\,,\quad\quad\bar{Q}_{\mu}:=Q{}_{\alpha}{}^{
\alpha}{}_{\mu}\,,\label{eq:Qvectors}
\end{align}
one has the relation  
\begin{align}
\mathring{R} & =Q+B\,,\label{eq:R_Q_B}
\end{align}
where $\mathring{R}$ is the Ricci scalar of the Levi-Civita connection, and 
  the boundary term is  
\begin{equation}
B:=\mathring{\nabla}_{\alpha}\left(\bar{Q}^{\alpha}-Q^{\alpha}\right).
\end{equation}
Therefore, General Relativity, which uses $\mathring{R}$ as the Lagrangian 
density, and STEGR, which uses $Q$, lead to exactly the same equations.

One can extend STEGR by extending the Lagrangian to an arbitrary function, 
resulting to $f(Q)$ gravity, with action 
\cite{BeltranJimenez:2017tkd,Harko:2018gxr} 
\begin{align}
S_{G} & ={\displaystyle \int 
d^{4}x\,\sqrt{-g}\left[-\frac{1}{16\pi G}f(Q)+\mathcal{L}_{m}  
\right]\,,}\label{eq:Action} 
\end{align}
where $G$ is the gravitational constant and $\mathcal{L}_{m}$ the matter 
Lagrangian.
The corresponding field equations are 
\begin{eqnarray}
&&\!\!\!\!\!\!\!\!\!\!\!\!\!\!\!\!\!\!\!\!\!\!\!
\sqrt{-g}
\left(\frac{1}{2}fg_{\mu\nu}-
\frac {\partial f}{\partial  g^{\mu\nu}}\right)
-2\nabla_{\alpha}\left(\sqrt{-g}P^{\alpha}{}_{\mu\nu}\right)  \nonumber\\
&&\ \ \ \ \ \ \ \ \ \ \ \ \ \ \ \ \ \ \ \ 
=8\pi G\sqrt{-g} \,T_{\mu\nu}\,,\label{eq:STG_field_eqn}
\end{eqnarray}
with 

\begin{align}
\!\!\!\!\!\!\!\!\!\!\!
\frac{\partial f}{\partial g^{\mu\nu}} & 
=-\frac{f_{\text{Q}}}{\sqrt{-g}}\left(\frac{\partial\left(\sqrt{-g}Q\right)}{
\partial g^{\mu\nu}}-\frac{1}{2}\sqrt{-g}Qg_{\mu\nu}\right)\nonumber \\
 & 
=-f_{\text{Q}}\biggl[c_{1}\left(2Q_{\alpha\beta\mu}Q^{\alpha\beta}{}_{\nu}-Q_{
\mu\alpha\beta}Q^{\nu}{}_{\alpha\beta}\right)\nonumber \\
 & \ \ \ \   \ \ \ \ \ \ \,
+c_{2}Q_{\alpha\beta\mu}Q^{\beta\alpha}{}_{\nu}+c_{3}\left(2Q_{\alpha}Q^{\alpha}
{}_{\mu\nu}\!-\!Q_{\mu}Q_{\nu}\right)\nonumber \\
 &  \ \ \ \   \ \ \ \ \ \ \,
+c_{4}\widetilde{Q}_{\mu}\widetilde{Q}_{\nu}+c_{5}\widetilde{Q}_{\mu}Q^{\alpha}{
}_{\mu\nu}\biggl],\label{eq:dfdg}
\end{align}
and
where the conjugate to $f(Q)$ is defined as
\cite{BeltranJimenez:2017tkd} 
\begin{align}
P^{\alpha}{}_{\mu\nu} & :=\frac{1}{2\sqrt{-g}}\frac{\partial(\sqrt{-g}f(Q))}{\partial Q{}_{\alpha}{}^{\mu\nu}}\,\nonumber \\
 & \phantom{:}=-f_{\text{Q}}\biggl\{c_{1}Q^{\alpha}{}_{\mu\nu}+c_{2}Q_{(\mu}{}^{\alpha}{}_{\nu)}+c_{3}g_{\mu\nu}Q^{\alpha}\nonumber \\
 & \quad\,+c_{4}\delta{}^{\alpha}{}_{(\mu}\widetilde{Q}_{\nu)}+\frac{c_{5}}{2}\bigl[\widetilde{Q}^{\alpha}g_{\mu\nu}+\delta{}^{\alpha}{}_{(\mu}Q_{\nu)}\bigr]\biggl\}\,,\label{eq:Ptensor}
\end{align}
while as usual  the matter energy momentum tensor is defined as 
\begin{align}
T_{\mu\nu} & :=-\frac{2}{\sqrt{-g}}\frac{\delta(\sqrt{-g}\mathcal{L}_{m})}{\delta g^{\mu\nu}}.\label{eq:Ttensor}
\end{align}
The field equations \eqref{eq:STG_field_eqn} are quite general with respect to 
both their normalizations described by $G$  and also the defined constants 
$c_i$ of $Q$. For instance, by choosing $Q$ as in Eq.~\eqref{eq: Q_GR} then we 
retrieve the standard modified class of $f(Q)$ gravity 
\cite{BeltranJimenez:2017tkd}. In the rest of this work we deal with this 
standard modified class of $f(Q)$ theories.

\section{Cosmological application of $f(Q)$ gravity}
 \label{sec:Cosmological-application-of}

 In this section we proceed to the cosmological application of $f(Q)$ gravity.  
We impose a spatially flat Friedmann-Robertson-Walker (FLRW) metric
of the form 
\begin{align}
ds^{2} & =-dt^{2}+a^{2}(t)\delta_{ij}dx^{i}dx^{j}\,,
\end{align}
with $a(t)$ the scale factor. In this  case Eq. 
\eqref{eq:STG_field_eqn}  
gives rise to the two Friedmann equations \cite{Jimenez:2019ovq}
\begin{align}
6f_{\text{Q}}H^{2}-\frac{1}{2}f & =8\pi G(\rho_{m}+\rho_{r})\,,\label{eq:Fried1}\\
\big(12H^{2}f_{\text{QQ}}+f_{\text{Q}}\big)\dot{H} & =-4\pi G(\rho_{m}+p_{m}+\rho_{r}+p_{r})\,,\label{eq:Fried2}
\end{align}
where $H=\dot{a}/a$ is the the Hubble function and with $\rho_{m}$,
$\rho_{r}$ and $p_{m}$, $p_{r}$ the energy densities and pressures
of the matter and radiation perfect fluids respectively. Additionally, note that
the non-metricity scalar $Q$ in an FRW background becomes $Q=6H^{2}$.
Finally, the equations constitute a close system by   considering
the matter and radiation conservation equations, namely
\begin{align}
\dot{\rho}_{m}+3H(\rho_{m}+p_{m}) & =0\,,\\
\dot{\rho}_{r}+3H(\rho_{r}+p_{r}) & =0\,.
\end{align}
We can re-write Eqs. \eqref{eq:Fried1}-\eqref{eq:Fried2} as 
\begin{align}
H^{2} & =\frac{8\pi G}{3}\left(\rho_{m}+\rho_{r}+\rho_{DE}\right)\,,\label{eq:Friedeff1}\\
\dot{H} & =-4\pi G\left(\rho_{m}+p_{m}+\rho_{r}+p_{r}+\rho_{DE}+p_{DE}\right)\,,\label{eq:Friedeff2}
\end{align}
where we have defined the energy density and pressure of the effective
dark energy sector as 
\begin{eqnarray}
&&\!\!\!\!\!\!\!\!\!\!\!\!\!\!\!\!\!\!\rho_{DE}  =\frac{1}{16\pi 
G}\left[6H^{2}\left(1-2f_{\text{Q}}\right)+f\right]\,,\label{eq:rhoDE}\\
\nonumber \\
&&\!\!\!\!\!\!\!
\!\!\!\!\!\!\!\!\!\!\!
p_{DE}  =\frac{1}{16\pi 
G}\left[4\left(f_{\text{Q}}-1\right) \dot{H}-f\right.\nonumber\\
&&\left.
\ \ \ \ \ \ \ \ \ \ 
+6H^{2}\left(8f_{\text{QQ}}\dot{H}+2f_{\text{Q}}
-1\right)\right]\,. \label{ eq:pDE }
\end{eqnarray}
Finally, it proves convenient to introduce the density parameters for the 
various sectors us
\begin{eqnarray}
 \Omega_i=\frac{8\pi G\rho_i}{3H^2},
\end{eqnarray}
where ``i'' stands for matter, radiation and dark energy.

Let us now focus on particular $f(Q)$ forms that are of interest for the late 
Universe description. In general, any model that exhibits 
identical Hubble rate with the concordance one (see \cite{Barros:2020bgg})  
can pass the BBN constraints trivially, hence we do not examine models of this 
kind. Note also that we consider versions of our models in their ``bare'' form, 
i.e. without explicitly containing the cosmological constant, in order to 
avoid re-introducing the cosmological constant problem and also to keep the 
smallest possible set of free parameters. One should in principle combine any 
selection of the following models, however the final model could have a large 
number of free parameters.

\begin{enumerate}
\item Polynomial model \\

The following model was introduced in \cite{Jimenez:2019ovq} 
\begin{align}
f(Q) & =Q-6\lambda M^{2}\left(\frac{Q}{6M^{2}}\right)^{\alpha}\,,\label{eq:model1}
\end{align}
where $\lambda$ and $\alpha$ are dimensionless parameters. The parameter
$M$ corresponds to a mass scale that should be of the order $\sqrt{\Lambda}$, 
with $\Lambda$ the standard cosmological constant.
The case of $\alpha=0$ gives STEGR plus a cosmological constant equal
to $6\lambda M^{2}$, while the case $\alpha=1$ corresponds to STEGR with
$G\rightarrow G/(1-\lambda)$. More generally, $\alpha>1$ is mostly relevant
to early Universe, while $\alpha<1$ is able to describe dark energy and thus it
is relevant to late Universe. Moreover, in the latter case the existence
of an asymptotic GR limit at early times is apparent.
Finally, note that the case $\alpha=-1$ has been
confronted with late universe observations in \cite{Ayuso:2020dcu}.

In this model for the effective dark-energy density (\ref{eq:rhoDE}) we obtain 
\begin{equation}
\rho_{\text{DE}}=\frac{1}{16\pi 
G}\left[6^{-\alpha}\Omega_{\text{F0}}H_{0}^{-2\alpha}Q ^{\alpha+1
}\right]\,,\label{eq:rhode1}
\end{equation}
where $\Omega_{F0} = 1 - \Omega_{m0} - \Omega_{r0}$, and with the subscript 
denoting the value of a quantity at present time. Calculating the first 
Friedmann  \eqref{eq:Fried1} at present time
we find that the free parameter  $\lambda$   can be eliminated in terms of 
$\alpha$ and $M$ as
\begin{equation}
\lambda=\frac{1}{12\alpha+6}\left(\Omega_{\text{F0}}H_{0}{}^{-2\alpha}M^{
2\alpha-2}\right)\,. \label{eq:param1}
\end{equation} 

\item Power - Exponential model\\ 
\label{enu:Power-Exponential-model--}

This model has proposed in \cite{Anagnostopoulos:2021ydo}
and has been proven to provide slightly better fits to observational data than the concordance
model \cite{Anagnostopoulos:2021ydo}. It is characterized by 
\begin{align}
f(Q) & =Qe^{\lambda\frac{Q_{0}}{Q}}\,.\label{eq:model2}
\end{align}
As it is apparent, in case   $\lambda=0$  GR without a cosmological constant is 
recovered. In the past, where the term $Q_{0}/Q$ decreases, since the Hubble 
function increases, the model at hand effectively reduces to the power-law 
model, thus alleviating the cosmological
constant problem. In a
sense, this model behaves as infinite number of different power-law
models that coexist, and at each moment of the cosmic history one
of them becomes dominant. Another advantage of the model is that
it has the same number of free parameters with the concordance one,
while additionally containing rich phenomenology. 

In this case (\ref{eq:rhoDE}) gives
\begin{equation}
\rho_{\text{DE}}=\frac{1}{16\pi G}\left[Q -e^{\frac{\lambda 
Q_{0}}{Q }}\left(Q -2\lambda Q_{0}\right)\right] \,.\label{eq:rhode2}
\end{equation}
The parameter $\lambda$ is expressed from the first Friedmann 
Eq.~\eqref{eq:Fried1}
at present  as  
\begin{equation}
\lambda=\frac{1}{2}+W\left(\frac{\Omega_{\text{F0}}-1}{2\sqrt{e}}\right)\,,\label{eq:param2}
\end{equation}
where $W$ is the Lambert function. 
 
\item Log-square-root model \\

Let us now propose a new $f(Q)$ model. In particular, following the approach of 
\cite{Bamba:2010wb},
we introduce a logarithmic-square-root $f(Q)$ model as:
\begin{align}
f(Q) & =Q+nQ_{0}\sqrt{\frac{Q}{\lambda 
Q_{0}}}\ln\left(\lambda\frac{Q_{0}}{Q}\right)\,,\label{eq:model5}
\end{align}
with $n$ and $\lambda>0$ the model parameters. The corresponding effective 
energy density    (\ref{eq:rhoDE}) gives  
\begin{align}
\rho_{\text{DE}}  & =\frac{1}{16\pi 
G}\Omega_{\text{F0}}\sqrt{Q_{0}Q }\,,\label{eq:rhode5}
\end{align}
and 
the first Friedmann equation at present leads to
\begin{align}
n & =\frac{\Omega_{\text{F0}}}{2}\sqrt{\lambda}\,.\label{eq:param5}
\end{align}

\item Hyperbolic tangent - power model \\
 
We consider the      hyperbolic tangent power model, in similar lines 
with \cite{Wu:2010av}, namely we choose
\begin{align}
f(Q) & =Q+\lambda 
Q_{0}\left(\frac{Q}{Q_{0}}\right){}^{n}\tanh\left(\frac{Q_{0}}{Q}\right)\,
.\label{eq:model6}
\end{align}
In this case we have  
\begin{eqnarray}
&&\ \ \rho_{\text{DE}}    =\frac{1}{16\pi 
G}Q_{0}^{1-n}Q ^{n-1}\nonumber\\ 
&&\ \ \ \ \ \ \
\cdot 
\left[(1\!-\!2n)Q \tanh\left(\!\frac{Q_{0}}{
Q  } \!\right)\!+\!2Q_ { 
0}\text{sech}^{2}\left(\!\frac{Q_{0}}{Q }\!\right)\right],\label{eq:rhode6}
\end{eqnarray}
while  
\begin{align}
n & 
=\frac{1}{2}\left[-\coth(1)\Omega_{\text{F0}}+1+4\text{csch}(2)\right]\,.
\label{eq:param6}
\end{align}

\item DGP-like $f(Q)$ model - I \\

In the recent work \cite{Ayuso:2021vtj}     two new $f(Q)$ models were 
introduced, by requiring the extra terms in the modified Friedman equation
to be of the form $\sim H$. It was argued that a term proportional to $Q$ is 
responsible for the $H^2$ term, thus adding a term proportional to 
$\sqrt{Q}$   leads to a term  $\sim H$ in the Friedman equation. 
These two models resemble the Dvali-Gabadadze-Porrati cosmology 
\cite{dvali20004d} at the background level, and this is where they
acquire their names from.
We mention here that this resemblance is only at the background level, since the perturbations 
in the usual DGP model and $f(Q)$ gravity are fundamentally  different, 
and hence the present scenario 
does not share the known problem of usual DGP model in 
fitting the perturbation-related data (such as LSS and
CMB temperature/polarization) \cite{Fairbairn:2005ue,fang2008challenges}.

The first model of this kind reads 
as
\begin{equation}
f(Q)=\alpha\sqrt{Q}\log Q+2\beta Q,\label{fqofmodelI}
\end{equation}
where $\alpha, \ \beta$ are free parameters. 
We mention that in the original parametrization the free parameter $\alpha$ 
is not dimensionless,  and we use 
the form \eqref{fqofmodelI} in order to maintain compatibility with the results 
of \cite{Ayuso:2021vtj} (although it would be more convenient to  impose a 
re-scaling $\alpha \rightarrow \sqrt{Q_{0}} \alpha$   to ensure that 
$\alpha$ is dimensionless). Nevertheless, we stress that  inside the logarithm 
one should explicitly include a constant value of dimension $H_0^{-2}$, and for 
consistency in the present work we do apply this modification.

In this case (\ref{eq:rhoDE}) gives
 \begin{eqnarray}
&&\rho_{\text{DE}}    =\frac{1}{16\pi 
G}\big[\sqrt{6}H_{0}\sqrt{Q 
}\left(2\beta+\Omega_{\text{F0}}-1\right) \nonumber\\
&& \ \ \ \ \ \ \ \ \ \ \ \ \ \ \ \ \ \ \,  -2\beta Q +Q 
\big],\label{fqofmodelI_rde}
\end{eqnarray}
while the first Friedmann equation gives
  \begin{eqnarray}
\alpha   
=-\sqrt{\frac{3}{2}}H_{0}\left(2\beta+\Omega_{\text{F0}}-1\right).
\label{fqofmodelI_rde_param}
\end{eqnarray}
 Finally, note that near  the current time, the term $\sim H$ ceases to be 
negligible in comparison to $H^2$, thus giving rise to new phenomenology. 
However, in the past this term is negligible and the model is very close to the 
concordance one. 

\item DGP-like $f(Q)$  model - II \\

The second model introduced in  \cite{Ayuso:2021vtj} reads as
\begin{equation}
\ \ \ \ \,\ f(Q)   = 
\frac{Q\sqrt{u(Q)}\left[\!\sqrt{u(Q)}\!-\!\sqrt{\gamma}\text{arctanh}
\left(\!\frac { \sqrt {u(Q)}}{\sqrt{\gamma}}\right)\!\right]}{8\pi 
G\sqrt{Qu(Q)}}\,,\label{fqofmodelII}
\end{equation}
where $u(Q) = \gamma + \beta^2 Q$, and $\beta$,$\gamma$ are free parameters, 
different than zero. The particular choice of $u$ is motivated from the 
previous case  and includes also a mixing between $H^2$ and 
$H$ terms  \cite{Ayuso:2021vtj}.
In this case we acquire
\begin{align}
\rho_{\text{DE}}  & =\frac{1}{16\pi 
G}\left[Q -\sqrt{Q }\sqrt{\gamma+\beta^{2}Q }\right]\,,
\label{fqofmodelII_rde}
\end{align}
while
\begin{align}
\gamma & 
=-6H_{0}{}^{2}\left(\beta-\Omega_{\text{F0}}+1\right)\left(\beta+\Omega_{\text{
F0}}-1\right)\,. \label{fqofmodelII_rde_param}
\end{align}
Similarly to the previous case, in order to maintain dimensional consistency 
within the square root, we include an $H_0^{-2}$ normalization (i.e. we apply 
the re-scaling $\gamma \rightarrow Q_{0} \gamma $).

\end{enumerate}

\section{Big Bang Nucleosynthesis constraints }
\label{Sec3}

In this Section  we review the Big Bang Nucleosynthesis (BBN) formalism 
following 
\cite{Bernstein:1988ad,Kolb:1990vq,Olive:1999ij,Cyburt:2015mya}. The BBN takes 
place during the radiation
era and thus the energy density of relativistic particles needs to
be taken into account, namely
\begin{align}
{\displaystyle \rho_{r}} & =\frac{\pi^{2}}{30}g_{*}T^{4}\,,
\end{align}
where 
\begin{equation}
g_{*}\sim10\,\label{eq:gstar}
\end{equation}
is the effective number of degrees of freedom and $T$ is the temperature
(for more details of the BBN framework used in this work see the appendix
of \cite{Barrow:2020kug}). The calculation of the neutron abundance
is realized by taking into account the protons-neutron conversion
rate 
\begin{equation}
\lambda_{pn}(T)=\lambda_{(n+\nu_{e}\to p+e^{-})}+\lambda_{(n+e^{+}\to p+\bar{\nu}_{e})}+\lambda_{(n\to p+e^{-}+\bar{\nu}_{e})}\,,
\end{equation}
and its inverse $\lambda_{np}(T)$, thus the total rate is 
\begin{equation}
\lambda_{tot}(T)=4A\,T^{3}(4!T^{2}+2\times3!{\cal Q}T+2!{\cal Q}^{2})\,,\label{Lambdafin}
\end{equation}
where ${\cal Q}=m_{n}-m_{p}=1.29\times10^{-3}$GeV is the neutro-proton
mass difference and $A=1.02\times10^{-11}$GeV$^{-4}$.

Regarding the primordial mass fraction of $^{4}He$, it can be estimated
\cite{Kolb:1990vq} as 
\begin{equation}
Y_{p}:=\lambda\,\frac{2x(t_{f})}{1+x(t_{f})}\,,\label{Yp}
\end{equation}
with $\lambda=e^{-(t_{n}-t_{f})/\tau}$, $t_{f}$ the freeze-out time
of the weak interactions, $t_{n}$ the corresponding freeze-out time
of nucleosynthesis, $\tau$ the neutron mean lifetime and $x(t_{f})=e^{-{\cal Q}/T(t_{f})}$
the neutron-to-proton equilibrium ratio. The role of the function
$\lambda(t_{f})$ is to account for the fraction of neutrons that
decay into protons during the time interval $t\in[t_{f},t_{n}]$.

In case of modified gravity models, in general the Friedmann equations
will contain extra terms than the standard GR ones. The BBN is realized in the 
radiation   epoch and according to observations
these extra contributions have to be small compared to the radiation
sector in the Standard Model of particles physics in the framework of General 
Relativity,  whilst we can safely neglect the matter sector as we 
are deep in the radiation era. Thus,
the first Friedmann equation can be approximated as 
\begin{align}
H^{2} & \approx\frac{8\pi G}{3}\rho_{r}=:H_{GR}^{2}\,,
\end{align}
where the scale factor evolves as $a\sim t^{1/2}$, with $t$ the
cosmic time. Consequently, temperature and time are related by ${\displaystyle \frac{1}{t}\simeq\left(\frac{32\pi^{3}g_{*}}{90}\right)^{1/2}\frac{T^{2}}{M_{P}}}$
(or $T(t)\simeq(t/\text{sec})^{-1/2}$MeV), which can further lead
to 
\begin{align}
H & \approx\left(\frac{4\pi^{3}g_{*}}{45}\right)^{1/2}\frac{T^{2}}{M_{P}}\,,
\end{align}
where 
\begin{equation}
M_{P}=(8\pi G)^{-1}=1.22\times10^{19}\text{GeV}\,,\label{eq:mp}
\end{equation}
is the Planck mass.

Assuming that the expansion time is much smaller than the interaction
time, then the interaction rate $\lambda_{tot}(T)$ given in (\ref{Lambdafin})
satisfies $\frac{1}{H}\ll\lambda_{tot}(T)$, which means that all processes
can be approximated as being in thermal equilibrium \cite{Kolb:1990vq,bernstein1989cosmological}.
In contrast, if the particles do not have the necessary time intervals
to interact then $\frac{1}{H}\gg\lambda_{tot}(T)$ and thus they decouple.
The temperature at which the particles decouple is called freeze-out
temperature and is denoted as $T_{f}$, and it is  defined
through $H=\lambda_{tot}(T)|_{T=T_f}$. Using 
$H\approx\left(\frac{4\pi^{3}g_{*}}{45}\right)^{1/2}\frac{T^{2}}{M_{P}}$,
while $\lambda_{tot}(T)\approx qT^{5}$, with 
\begin{equation}
q=4A4!\simeq9.8\times10^{-10}\text{GeV}{}^{-4}\,,\label{eq:qvalue}
\end{equation}
the freeze-out temperature is provided as 
\begin{align}
T_{f} & =\left(\frac{4\pi^{3}g_{*}}{45M_{P}^{2}q^{2}}\right)^{1/6}\sim0.0006\,\text{GeV}\,.\label{Tfreeze}
\end{align}

In the realm of modified gravity, the Hubble function $H$ will, in
general, deviate from $H_{GR}$. This in turn means that the corresponding
freeze-out temperatures will also deviate by an amount of $\delta T_{f}$.
This deviation induces a difference $\delta Y_{p}$ of the fractional
mass $Y_{p}$ 
\begin{align}
\delta Y_{p} & =Y_{p}\left[\left(1-\frac{Y_{p}}{2\lambda}\right)\ln\left(\frac{2\lambda}{Y_{p}}-1\right)-\frac{2t_{f}}{\tau}\right]\frac{\delta T_{f}}{T_{f}}\,,\label{deltaYp}
\end{align}
where $\delta T(t_{n})=0$ was imposed due to the fact that $T_{n}$
is fixed by the binding energy of deuterium \cite{Torres:1997sn,Lambiase:2005kb,Lambiase:2012fv,Lambiase:2011zz,Capozziello:2017bxm}.
Thus, the observational imprints of the mass fraction $Y_{p}$
of baryons that  convert to $^{4}He$ during the BBN epoch, are
\cite{Coc:2003ce,Olive:1996zu,Izotov:1998mj,Fields:1998gv,Izotov:1999wa,Kirkman:2003uv,Izotov:2003xn}
\begin{align}
Y_{p} & =0.2476\,,\qquad|\delta Y_{p}|<10^{-4}\,.\label{Ypoohdges}
\end{align}
By replacing these into \eqref{deltaYp}, the upper bound of $\frac{\delta 
T_{f}}{T_{f}}$
is obtained as 
\begin{align}
\left|\frac{\delta T_{f}}{T_{f}}\right| & <4.7\times10^{-4}\,,\label{deltaT/Tbound}
\end{align}
which quantifies the allowed deviation from the cosmology of GR.

As we mentioned above,  the effective dark energy $\rho_{DE}$  of modified 
gravity   is present during the BBN times too, and compared to $\rho_{r}$,
$\rho_{DE}$ should be smaller and thus it can be considered as a
first-order deviation. Hence, the Hubble function is 
\begin{align}
H & =H_{GR}\sqrt{1+\frac{\rho_{DE}}{\rho_{r}}}=H_{GR}+\delta H,\label{H11}
\end{align}
where 
\begin{align}
\delta H & =\left(\sqrt{1+\frac{\rho_{DE}}{\rho_{r}}}-1\right)H_{GR}\,.\label{deltaH}
\end{align}
The deviation $\delta H$ from standard $H_{GR}$ will induce the
deviation $\delta T_{f}$  from $T_{f}$,  and according to the relation
$H_{GR}=\lambda_{tot}\approx qT^{5}$ it turns out that 
\begin{equation}
\left(\sqrt{1+\frac{\rho_{DE}}{\rho_{r}}}-1\right)H_{GR}=5qT_{f}^{4}\delta 
T_{f}\,.\label{H_T=00003D00003D00003DLambda}
\end{equation}
Since $\rho_{DE}\ll\rho_{r}$ we finally obtain
\begin{equation}
\frac{\delta T_{f}}{T_{f}}\simeq\frac{\rho_{DE}}{\rho_{r}}\frac{H_{GR}}{10qT_{f}^{5}}\,.\label{deltaTfcon11}
\end{equation}

We have now all the information needed to proceed to the investigation
of the BBN bounds on the parameters of the $f(Q)$ models introduced above.

\section{BBN constraints on $f(Q)$ gravity}
\label{Sec4}

In this section, the models introduced in 
Sec.~\ref{sec:Cosmological-application-of}
will be tested against the constraint \eqref{deltaT/Tbound}. This
will be realized by calculating  \eqref{deltaTfcon11} for each
of the models. In general, we are interested in the regions  of the 
parameter
space where the constraint \eqref{deltaTfcon11} is satisfied. Obviously,
all models are expected to satisfy the constraint trivially for the
case $\Omega_{\text{F0}}=0$, i.e.  in the case where there is no effective 
dark energy. In the following, in all figures the 
drawn parts correspond to satisfaction of  the aforementioned BBN constraint.
For the standard free parameters we impose the ranges $h\in[0.6,0.9]$ and
$\Omega_{\text{F0}}\in[0,1]$. For the case of free model parameters
we use the mathematically allowed ranges.  

\subsection*{1. Polynomial model\label{subsec:1.Polynomial-model}}

\begin{figure}[H]
\hspace{-2.2cm} \includegraphics[scale=0.43]{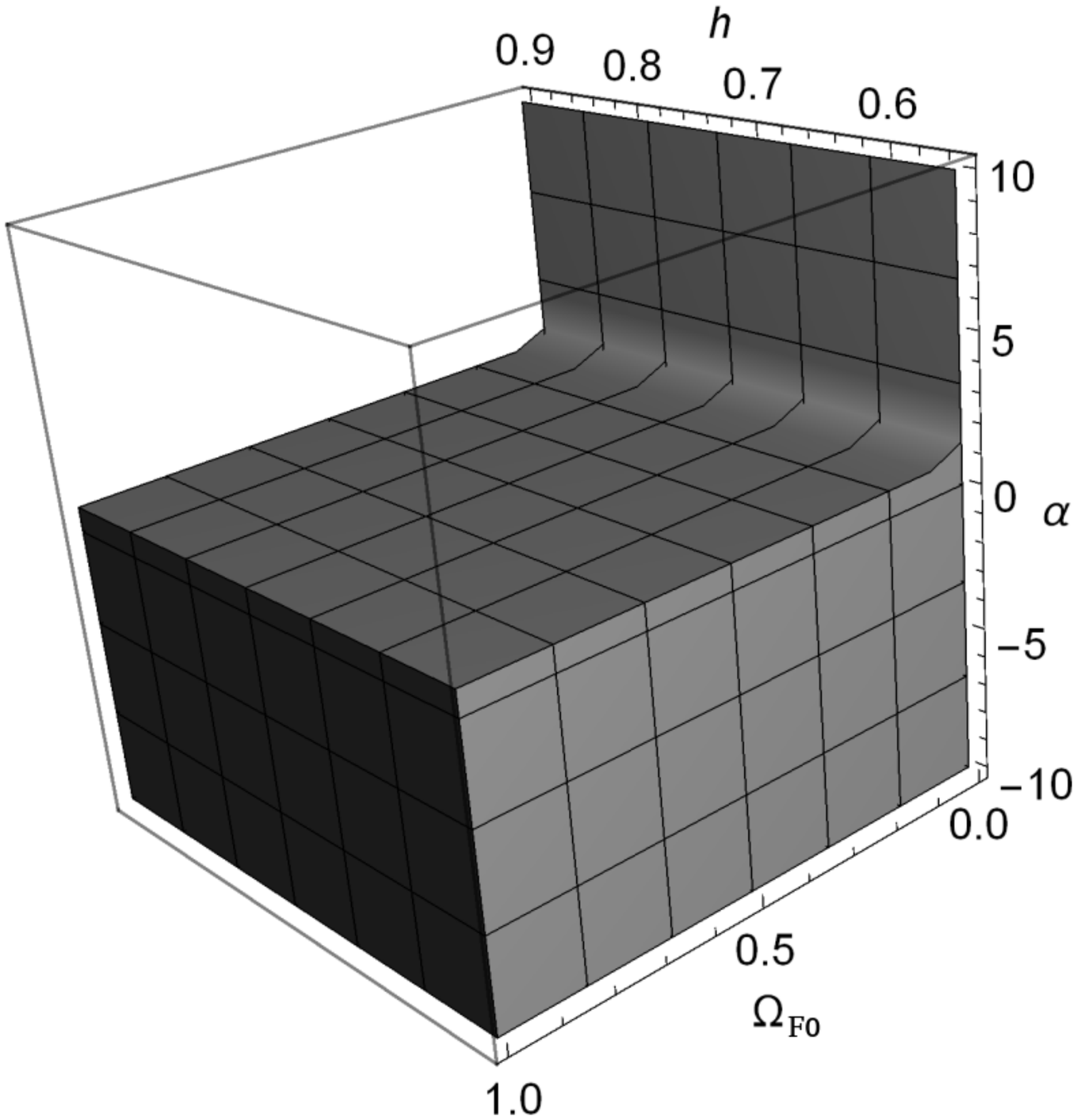} 
\caption[adf]{{\it{The parameter space 
 of the polynomial model \eqref{eq:model1} that is consistent with the BBN 
constraints. If we also desire the model 
to be able to describe dark energy in 
the late-time Universe we obtain the combined constraint $\alpha <0$.  }} }
\label{fig:model1} 
\end{figure}

Substituting Eq.~\eqref{eq:param1} along with Eq.~\eqref{eq:rhode1}
in Eq.~\eqref{deltaTfcon11} we obtain 
\begin{align}
\frac{\delta T_{f}}{T_{f}} & \simeq\frac{1}{q}2^{4\alpha+3}3^{\frac{1}{2}-\alpha}5^{-\alpha-\frac{3}{2}}\pi^{3\alpha+2}H_{0}\Omega_{\text{F0}}\text{}\nonumber \\
 & 
\phantom{\simeq}\cdot 
g_{*}^{\alpha+\frac{1}{2}}T_{f}^{4\alpha-3}\left(H_{0}M_{p} 
\right){}^{-2\alpha-1}\,.\label{eq:dt1}
\end{align}
Imposing the constraint \eqref{deltaT/Tbound} on the parameter space
of the polynomial model using  \eqref{eq:dt1}, we obtain 
Fig.~\ref{fig:model1}. Note that    we imposed the  relation
$H_{0}=2.12 \cdot 10^{-42}h$ GeV
and we have also replaced the numerical values \eqref{eq:qvalue},
\eqref{eq:gstar}, \eqref{Tfreeze} and \eqref{eq:mp}.  

From Fig. ~\ref{fig:model1} we deduce that the model parameter $\alpha$ 
should be in the range  $\alpha \lesssim 0.88$ in order for the model to satisfy the 
BBN constraints. However,  if we additionally desire the model to be able to 
describe dark energy in the late-time  Universe, then we obtain the combined 
constraint  $\alpha <0$.

\subsection*{2. Power-Exponential model}

\label{enu:Power-Exponential-model---2}

Substituting Eq.~\eqref{eq:param2} alongside with  \eqref{eq:rhode2}
into  \eqref{deltaTfcon11} we acquire 
\begin{equation}
\frac{\delta T_{f}}{T_{f}}\simeq\frac{8\sqrt{\frac{3}{5}}\pi^{2}\sqrt{g}T_{0}^{4}\left\{ e^{\lambda\left(\frac{T_{0}}{T_{f}}\right)^{4}}\left[2\lambda-\left(\frac{T_{f}}{T_{0}}\right)^{4}\right]+\left(\frac{T_{f}}{T_{0}}\right)^{4}\right\} }{5qT_{f}^{7}M_{p}}\,.\label{eq:dt2}
\end{equation}
Inserting this expression  into \eqref{deltaT/Tbound} we 
find 
that for the range of values $0.5\leq h\leq0.9$ and $0\leq\Omega_{\text{F0}}\leq1$,
the constraint  \eqref{deltaT/Tbound} is satisfied trivially for all parameter 
values. This result is intuitively expected, since as we mentioned above the 
past asymptotic behavior of the model recovers pure GR at early times, and 
thus during the BBN epoch 
\cite{Anagnostopoulos:2021ydo}. This is a great advantage of this particular 
model.

\subsection*{3. Log-square-root}

\label{enu:Log-square-root-model-1}

In this case, inserting \eqref{eq:param5} and \eqref{eq:rhode5}
in \eqref{deltaTfcon11} we find 
\begin{equation}
\frac{\delta T_{f}}{T_{f}}\simeq8\pi^{2}\sqrt{\frac{3g}{5}}\frac{\Omega_{\text{F0}}T_{0}^{2}}{5qT_{f}^{5}M_{p}}\,.\label{eq:dt5}
\end{equation}
Therefore, substituting \eqref{eq:dt5} in \eqref{deltaT/Tbound} we observe
that for the range of values $0.5\leq h\leq0.9$ and $0\leq\Omega_{\text{F0}}\leq1$,
the constraint  \eqref{deltaT/Tbound} is always satisfied. This is an advantage 
of the model, and supports its viability. Definitely, one should   confront 
the model  with late-time 
observational data, too.

\subsection*{4. Hyperbolic tangent - power model}
\label{enu:Trigonometric-model}

\begin{figure}[H]
\vspace{1cm}
\hspace{-2.2cm} \includegraphics[scale=0.43]{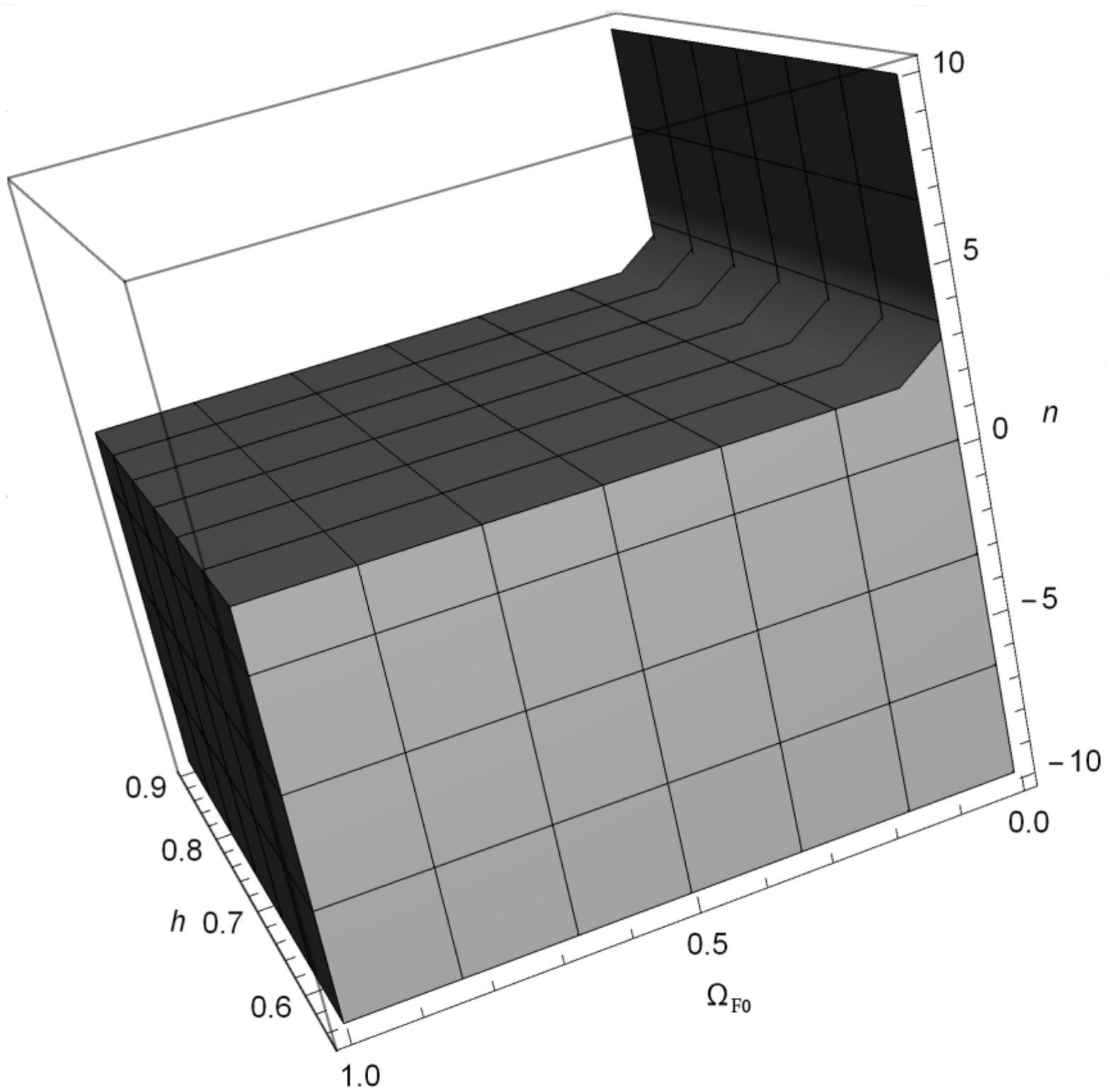} \caption[adf]{
{\it{The parameter space 
 of the hyperbolic tangent - power model \eqref{eq:model6} that is consistent 
with the BBN 
constraints. If we also desire the model 
to be able to describe dark energy in 
the late-time Universe we obtain the combined constraint $n\lesssim1.88$.  }} 
 }
\label{fig:model6} 
\end{figure}

In the case of the hyperbolic tangent - power model (\ref{eq:model6}),
substituting  \eqref{eq:param6} and \eqref{eq:rhode6}
into \eqref{deltaTfcon11} we obtain
 \begin{eqnarray}
&&\!\!\!\!\!\!\frac{\delta T_{f}}{T_{f}}    
\simeq\sqrt{\frac{3}{5}}\frac{8\pi^{2}\sqrt{g}}{5qM_{p}T_{f}^{7}} 
\left(\frac{T_{f}}{T_{0}}\right)^{4n-4}\nonumber\\
&&\!\!\!\!\!\!\cdot
 \left\{ \!
(1\!-\!2n)T_{f}^{4}\tanh\!\left[\!\left(\frac{T_{0}}{T_{f}}\right)^{4}\right]
\!+\!2T_{
0} ^{ 4 } 
\text{sech}^{2}\!\left[\!\left(\frac{T_{0}}{T_{f}}\right)^{4}\right]\!\right\} 
 .\label{eq:dt6}
\end{eqnarray}
 In Fig.~\ref{fig:model6} we depict  the constraint \eqref{deltaT/Tbound} by 
using  \eqref{eq:dt6}. The procedure is the same
as in case of Fig. \ref{fig:model1} above. As we observe,  the BBN constraints 
are satisfied for   $n\lesssim1.88$.

\subsection*{5. DGP-like model - I}

For the DGP-like model - I  \eqref{fqofmodelI}, if we insert 
 \eqref{fqofmodelI_rde_param} and \eqref{fqofmodelI_rde}
into \eqref{deltaTfcon11}, we acquire
\begin{equation}
\frac{\delta T_{f}}{T_{f}}   
\simeq\frac{8\sqrt{\frac{3}{5}}\pi^{2}\sqrt{g}T_{0}{}^{2}\left[
2\beta\!+\!\Omega_{ \text{F0}}\!-\!2\beta\left(\frac{T_{f}}{T_{0}}\right)^{2}
\!+\!\left(\frac{T_{f}}{T_{0}}
\right)^{2}\!-\!1\right]}{5qM_{p}T_{f}{}^{5}} .
\end{equation}
Furthermore, numerical exploration of this equation for the standard parameter 
ranges  $0.5\leq h\leq0.9$ and $0\leq\Omega_{\text{F0}}\leq1$, shows 
that
the constraint  \eqref{deltaT/Tbound} is satisfied only in the tuned interval
   $0.499984 < \beta < 0.500016$.

\subsection*{6. DGP-like model - II}

Finally, let us come to the DGP-like model - II of \eqref{fqofmodelII}.
The constraint \eqref{deltaTfcon11}, after inserting 
 \eqref{fqofmodelII_rde_param} and \eqref{fqofmodelII_rde}, 
reduces to
\begin{eqnarray}
&&
\!\!\!\!\!
\frac{\delta T_{f}}{T_{f}}   \simeq
\sqrt{\frac{3}{5}}\frac{8\pi^{2}\sqrt{g}T_{0}{}^{2}}{5qM_{p}T_{f}^{5}} 
\nonumber\\
&&\!\!\!\!\!\cdot 
\left\{\!\!\left(\frac{T_{f}}{T_{0}}\right)^{2}\!\!-\!\sqrt{\!\left(\Omega_{
\text{ F0 }} \!
-\!2\right)\!\Omega_{\text{F0}}\!+\!\beta^{2}\!\left[\!\left(\frac{T_{f}}{T_{0}}
\right)^ {4 } \!-\!1\!\right]\!+\!1}\right\}  .
\end{eqnarray}
Examining this expression  for the range of values $0.5\leq h\leq0.9$ and 
$0\leq\Omega_{\text{F0}}\leq1$, we deduce that the constraint  
 \eqref{deltaT/Tbound} is satisfied in the tuned window $0.999969 < |\beta| < 
1.00003$.

\section{Conclusions}
\label{Conclusions}

We use the Big Bang Nucleosynthesis formalism and observations in order to 
extract constraints on $f(Q)$ gravity. The latter is a modified gravitational 
theory arising from the consideration of non-metricity as the basic geometric 
quantity. We investigated various classes of $f(Q)$ models, and we additionally 
introduced a number of new ones, inspired by other modified gravities, that are 
capable of describing the late-time Universe evolution. We  
applied the semi-analytic approach of 
\cite{Lambiase:2005kb,Lambiase:2012fv,Lambiase:2011zz} for the physics of 
the  BBN epoch, and  we 
calculated the deviations that $f(Q)$ terms bring on the freeze-out 
temperature $T_f$ in  comparison  to that of the standard 
$\Lambda$CDM evolution. We then imposed the observational  bound on $ 
\left|\frac{\delta {T}_f}{{T}_f}\right|$   to extract constraints on the 
involved parameters of the considered  models.

Concerning the  polynomial model, we showed that the exponent parameter 
$\alpha$ should be in the interval $\alpha <0$ if we desire the model to 
simultaneously pass the BBN constraints and be able to describe the late-time 
Universe acceleration. On the other hand, the power-exponential model is found 
to pass the BBN constraints trivially, which was expected since this model 
recovers general relativity at early times. This is a significant advantage, 
since this particular model is known to be compatible with late-time 
observational datasets slightly more efficiently than $\Lambda$CDM scenario 
\cite{Anagnostopoulos:2021ydo}, and thus it can be considered as a good 
candidate for the description of Nature. 

In the case of  the new proposed Log-square-root models we showed that BBN 
constraints are also trivially satisfied. However, in the new    hyperbolic 
tangent - power model we showed that the combined BBN and late-time constraints
require the model parameter to lie in the range $n \lesssim 1.88$. Finally, we 
examined the two    DGP-like $f(Q)$
models that have recently appeared in the literature. In both cases we found 
that the BBN constraints are satisfied for a very narrow window for the model 
parameter $\beta$.

In summary, we showed that $f(Q)$ gravity can safely pass the BBN constraints, 
and in some cases this is obtained trivially. This is an important advantage, 
since many gravitational modifications, although able to describe the late-time 
evolution of the Universe, fall to pass the BBN confrontation since they 
produce too-much modification at early times. Hence, $f(Q)$ gravity proves to 
be a class of modified gravity that deserves further investigation. One could 
try to proceed   beyond the 
semi-analytical approach of this work, using numerical codes like Parthenope  
\cite{gariazzo2021parthenope} and AlterBBN  \cite{Arbey:2018zfh}. Such a 
detailed investigation lies beyond the scope of the present work and it is left 
for a future project.\\

\begin{acknowledgments}
 
The authors   acknowledge participation in the COST Association 
Action CA18108 ``{\it Quantum Gravity Phenomenology in the Multimessenger 
Approach (QG-MM)}''.

\end{acknowledgments}

 \bibliographystyle{utphys}
\bibliography{references}

\providecommand{\href}[2]{#2}\begingroup\raggedright\begin{thebibliography}{10}

\bibitem{Aghanim:2018eyx}
{\bf Planck} Collaboration, N.~Aghanim {\em et al.}, ``{Planck 2018 results.
  VI. Cosmological parameters},'' \href{http://arxiv.org/abs/1807.06209}{{\tt
  arXiv:1807.06209 [astro-ph.CO]}}.

\bibitem{scolnic2018complete}
D.~M. Scolnic, D.~Jones, A.~Rest, Y.~Pan, R.~Chornock, R.~Foley, M.~Huber,
  R.~Kessler, G.~Narayan, A.~Riess, {\em et al.}, ``The complete light-curve
  sample of spectroscopically confirmed sne ia from pan-starrs1 and
  cosmological constraints from the combined pantheon sample,'' {\em The
  Astrophysical Journal} {\bf 859} (2018) no.~2, 101. The numerical data of the
  full Pantheon SnIa sample are available at
  \url{http://dx.doi.org/10.17909/T95Q4X},
  \url{https://archive.stsci.edu/prepds/ps1cosmo/index.html.}

\bibitem{Gil-Marin:2015sqa}
H.~Gil-Mar\'\i{}n {\em et al.}, ``{The clustering of galaxies in the SDSS-III
  Baryon Oscillation Spectroscopic Survey: RSD measurement from the
  LOS-dependent power spectrum of DR12 BOSS galaxies},''
  \href{http://dx.doi.org/10.1093/mnras/stw1096}{{\em Mon. Not. Roy. Astron.
  Soc.} {\bf 460} (2016) no.~4, 4188--4209},
  \href{http://arxiv.org/abs/1509.06386}{{\tt arXiv:1509.06386 [astro-ph.CO]}}.

\bibitem{di2020cosmology}
E.~Di~Valentino, L.~A. Anchordoqui, O.~Akarsu, Y.~Ali-Haimoud, L.~Amendola,
  N.~Arendse, M.~Asgari, M.~Ballardini, S.~Basilakos, E.~Battistelli, {\em et
  al.}, ``Cosmology intertwined ii: the hubble constant tension,'' {\em arXiv
  preprint arXiv:2008.11284} (2020)  .

\bibitem{di2021cosmology}
E.~Di~Valentino, L.~A. Anchordoqui, {\"O}.~Akarsu, Y.~Ali-Haimoud, L.~Amendola,
  N.~Arendse, M.~Asgari, M.~Ballardini, S.~Basilakos, E.~Battistelli, {\em et
  al.}, ``Cosmology intertwined iii: f$\sigma$8 and s8,'' {\em Astroparticle
  Physics} {\bf 131} (2021)  102604.

\bibitem{weinberg1989cosmological}
S.~Weinberg, ``The cosmological constant problem,'' {\em Reviews of modern
  physics} {\bf 61} (1989) no.~1, 1.

\bibitem{Addazi:2021xuf}
A.~Addazi {\em et al.}, ``{Quantum gravity phenomenology at the dawn of the
  multi-messenger era\textemdash{}A review},''
  \href{http://dx.doi.org/10.1016/j.ppnp.2022.103948}{{\em Prog. Part. Nucl.
  Phys.} {\bf 125} (2022)  103948}, \href{http://arxiv.org/abs/2111.05659}{{\tt
  arXiv:2111.05659 [hep-ph]}}.

\bibitem{stelle1977renormalization}
K.~S. Stelle, ``Renormalization of higher-derivative quantum gravity,'' {\em
  Physical Review D} {\bf 16} (1977) no.~4, 953.

\bibitem{CANTATA:2021ktz}
{\bf CANTATA} Collaboration, E.~N. Saridakis {\em et al.}, ``{Modified Gravity
  and Cosmology: An Update by the CANTATA Network},''
  \href{http://arxiv.org/abs/2105.12582}{{\tt arXiv:2105.12582 [gr-qc]}}.

\bibitem{Capozziello:2011et}
S.~Capozziello and M.~De~Laurentis, ``{Extended Theories of Gravity},''
  \href{http://dx.doi.org/10.1016/j.physrep.2011.09.003}{{\em Phys. Rept.} {\bf
  509} (2011)  167--321},
\href{http://arxiv.org/abs/1108.6266}{{\tt arXiv:1108.6266 [gr-qc]}}.

\bibitem{Abdalla:2022yfr}
E.~Abdalla {\em et al.}, ``{Cosmology intertwined: A review of the particle
  physics, astrophysics, and cosmology associated with the cosmological
  tensions and anomalies},''
  \href{http://dx.doi.org/10.1016/j.jheap.2022.04.002}{{\em JHEAp} {\bf 34}
  (2022)  49--211}, \href{http://arxiv.org/abs/2203.06142}{{\tt
  arXiv:2203.06142 [astro-ph.CO]}}.

\bibitem{DeFelice:2010aj}
A.~De~Felice and S.~Tsujikawa, ``{f(R) theories},''
  \href{http://dx.doi.org/10.12942/lrr-2010-3}{{\em Living Rev. Rel.} {\bf 13}
  (2010)  3},
\href{http://arxiv.org/abs/1002.4928}{{\tt arXiv:1002.4928 [gr-qc]}}.

\bibitem{Antoniadis:1993jc}
I.~Antoniadis, J.~Rizos, and K.~Tamvakis, ``{Singularity - free cosmological
  solutions of the superstring effective action},''
  \href{http://dx.doi.org/10.1016/0550-3213(94)90120-1}{{\em Nucl. Phys. B}
  {\bf 415} (1994)  497--514}, \href{http://arxiv.org/abs/hep-th/9305025}{{\tt
  arXiv:hep-th/9305025}}.

\bibitem{Nojiri:2005jg}
S.~Nojiri and S.~D. Odintsov, ``{Modified Gauss-Bonnet theory as gravitational
  alternative for dark energy},''
  \href{http://dx.doi.org/10.1016/j.physletb.2005.10.010}{{\em Phys. Lett. B}
  {\bf 631} (2005)  1--6}, \href{http://arxiv.org/abs/hep-th/0508049}{{\tt
  arXiv:hep-th/0508049}}.

\bibitem{Mannheim:1988dj}
P.~D. Mannheim and D.~Kazanas, ``{Exact Vacuum Solution to Conformal Weyl
  Gravity and Galactic Rotation Curves},''
\href{http://dx.doi.org/10.1086/167623}{{\em Astrophys. J.} {\bf 342} (1989)
  635--638}.

\bibitem{Bengochea:2008gz}
G.~R. Bengochea and R.~Ferraro, ``{Dark torsion as the cosmic speed-up},''
  \href{http://dx.doi.org/10.1103/PhysRevD.79.124019}{{\em Phys. Rev.} {\bf
  D79} (2009)  124019},
\href{http://arxiv.org/abs/0812.1205}{{\tt arXiv:0812.1205 [astro-ph]}}.

\bibitem{Cai:2015emx}
Y.-F. Cai, S.~Capozziello, M.~De~Laurentis, and E.~N. Saridakis, ``{f(T)
  teleparallel gravity and cosmology},''
  \href{http://dx.doi.org/10.1088/0034-4885/79/10/106901}{{\em Rept. Prog.
  Phys.} {\bf 79} (2016) no.~10, 106901},
\href{http://arxiv.org/abs/1511.07586}{{\tt arXiv:1511.07586 [gr-qc]}}.

\bibitem{Kofinas:2014owa}
G.~Kofinas and E.~N. Saridakis, ``{Teleparallel equivalent of Gauss-Bonnet
  gravity and its modifications},''
  \href{http://dx.doi.org/10.1103/PhysRevD.90.084044}{{\em Phys. Rev.} {\bf
  D90} (2014)  084044},
\href{http://arxiv.org/abs/1404.2249}{{\tt arXiv:1404.2249 [gr-qc]}}.

\bibitem{Geng:2011aj}
C.-Q. Geng, C.-C. Lee, E.~N. Saridakis, and Y.-P. Wu, ``{“Teleparallel”
  dark energy},'' \href{http://dx.doi.org/10.1016/j.physletb.2011.09.082}{{\em
  Phys. Lett.} {\bf B704} (2011)  384--387},
\href{http://arxiv.org/abs/1109.1092}{{\tt arXiv:1109.1092 [hep-th]}}.

\bibitem{Hohmann:2018rwf}
M.~Hohmann, L.~J{\"a}rv, and U.~Ualikhanova, ``{Covariant formulation of
  scalar-torsion gravity},''
  \href{http://dx.doi.org/10.1103/PhysRevD.97.104011}{{\em Phys. Rev.} {\bf
  D97} (2018) no.~10, 104011},
\href{http://arxiv.org/abs/1801.05786}{{\tt arXiv:1801.05786 [gr-qc]}}.

\bibitem{BeltranJimenez:2017tkd}
J.~Beltr\'an~Jim\'enez, L.~Heisenberg, and T.~Koivisto, ``{Coincident General
  Relativity},'' \href{http://dx.doi.org/10.1103/PhysRevD.98.044048}{{\em Phys.
  Rev. D} {\bf 98} (2018) no.~4, 044048},
  \href{http://arxiv.org/abs/1710.03116}{{\tt arXiv:1710.03116 [gr-qc]}}.

\bibitem{Jimenez:2019ovq}
J.~Beltr\'an~Jim\'enez, L.~Heisenberg, T.~S. Koivisto, and S.~Pekar,
  ``{Cosmology in $f(Q)$ geometry},''
  \href{http://dx.doi.org/10.1103/PhysRevD.101.103507}{{\em Phys. Rev. D} {\bf
  101} (2020) no.~10, 103507}, \href{http://arxiv.org/abs/1906.10027}{{\tt
  arXiv:1906.10027 [gr-qc]}}.

\bibitem{Xu:2019sbp}
Y.~Xu, G.~Li, T.~Harko, and S.-D. Liang, ``{$f(Q,T)$ gravity},''
  \href{http://dx.doi.org/10.1140/epjc/s10052-019-7207-4}{{\em Eur. Phys. J. C}
  {\bf 79} (2019) no.~8, 708}, \href{http://arxiv.org/abs/1908.04760}{{\tt
  arXiv:1908.04760 [gr-qc]}}.

\bibitem{Jimenez:2019ghw}
J.~B. Jiménez, L.~Heisenberg, D.~Iosifidis, A.~Jiménez-Cano, and T.~S.
  Koivisto, ``{General Teleparallel Quadratic Gravity},''
\href{http://arxiv.org/abs/1909.09045}{{\tt arXiv:1909.09045 [gr-qc]}}.

\bibitem{Percacci:2019hxn}
R.~Percacci and E.~Sezgin, ``{New class of ghost- and tachyon-free metric
  affine gravities},''
  \href{http://dx.doi.org/10.1103/PhysRevD.101.084040}{{\em Phys. Rev. D} {\bf
  101} (2020) no.~8, 084040}, \href{http://arxiv.org/abs/1912.01023}{{\tt
  arXiv:1912.01023 [hep-th]}}.

\bibitem{Delhom:2020vpe}
A.~Delhom, I.~P. Lobo, G.~J. Olmo, and C.~Romero, ``{Conformally invariant
  proper time with general non-metricity},''
  \href{http://dx.doi.org/10.1140/epjc/s10052-020-7974-y}{{\em Eur. Phys. J. C}
  {\bf 80} (2020) no.~5, 415}, \href{http://arxiv.org/abs/2001.10633}{{\tt
  arXiv:2001.10633 [gr-qc]}}.

\bibitem{Delhom:2020hkb}
A.~Delhom, ``{Minimal coupling in presence of non-metricity and torsion},''
  \href{http://dx.doi.org/10.1140/epjc/s10052-020-8330-y}{{\em Eur. Phys. J. C}
  {\bf 80} (2020) no.~8, 728}, \href{http://arxiv.org/abs/2002.02404}{{\tt
  arXiv:2002.02404 [gr-qc]}}.

\bibitem{Barros:2020bgg}
B.~J. Barros, T.~Barreiro, T.~Koivisto, and N.~J. Nunes, ``{Testing $F(Q)$
  gravity with redshift space distortions},''
  \href{http://dx.doi.org/10.1016/j.dark.2020.100616}{{\em Phys. Dark Univ.}
  {\bf 30} (2020)  100616}, \href{http://arxiv.org/abs/2004.07867}{{\tt
  arXiv:2004.07867 [gr-qc]}}.

\bibitem{BeltranJimenez:2020sih}
J.~Beltr\'an~Jim\'enez, L.~Heisenberg, and T.~Koivisto, ``{The coupling of
  matter and spacetime geometry},''
  \href{http://dx.doi.org/10.1088/1361-6382/aba31b}{{\em Class. Quant. Grav.}
  {\bf 37} (2020) no.~19, 195013}, \href{http://arxiv.org/abs/2004.04606}{{\tt
  arXiv:2004.04606 [hep-th]}}.

\bibitem{Jimenez-Cano:2020chm}
A.~Jim\'enez-Cano, ``{New metric-affine generalizations of gravitational wave
  geometries},'' \href{http://dx.doi.org/10.1140/epjc/s10052-020-8239-5}{{\em
  Eur. Phys. J. C} {\bf 80} (2020) no.~7, 672},
  \href{http://arxiv.org/abs/2005.02014}{{\tt arXiv:2005.02014 [gr-qc]}}.

\bibitem{DAmbrosio:2020nev}
F.~D'Ambrosio, M.~Garg, and L.~Heisenberg, ``{Non-linear extension of
  non-metricity scalar for MOND},''
  \href{http://dx.doi.org/10.1016/j.physletb.2020.135970}{{\em Phys. Lett. B}
  {\bf 811} (2020)  135970}, \href{http://arxiv.org/abs/2004.00888}{{\tt
  arXiv:2004.00888 [gr-qc]}}.

\bibitem{Rubiera-Garcia:2020gcl}
D.~Rubiera-Garcia, ``{From fundamental physics to tests with compact objects in
  metric-affine theories of gravity},''
  \href{http://dx.doi.org/10.1142/S0218271820410072}{{\em Int. J. Mod. Phys. D}
  {\bf 29} (2020) no.~11, 2041007}, \href{http://arxiv.org/abs/2004.00943}{{\tt
  arXiv:2004.00943 [gr-qc]}}.

\bibitem{Xu:2020yeg}
Y.~Xu, T.~Harko, S.~Shahidi, and S.-D. Liang, ``{Weyl type $f(Q,T)$ gravity,
  and its cosmological implications},''
  \href{http://dx.doi.org/10.1140/epjc/s10052-020-8023-6}{{\em Eur. Phys. J. C}
  {\bf 80} (2020) no.~5, 449}, \href{http://arxiv.org/abs/2005.04025}{{\tt
  arXiv:2005.04025 [gr-qc]}}.

\bibitem{Ayuso:2020dcu}
I.~Ayuso, R.~Lazkoz, and V.~Salzano, ``{Observational constraints on
  cosmological solutions of $f(Q)$ theories},''
  \href{http://dx.doi.org/10.1103/PhysRevD.103.063505}{{\em Phys. Rev. D} {\bf
  103} (2021) no.~6, 063505}, \href{http://arxiv.org/abs/2012.00046}{{\tt
  arXiv:2012.00046 [astro-ph.CO]}}.

\bibitem{Cabral:2020fax}
F.~Cabral, F.~S.~N. Lobo, and D.~Rubiera-Garcia, ``{Fundamental Symmetries and
  Spacetime Geometries in Gauge Theories of Gravity\textemdash{}Prospects for
  Unified Field Theories},''
  \href{http://dx.doi.org/10.3390/universe6120238}{{\em Universe} {\bf 6}
  (2020) no.~12, 238}, \href{http://arxiv.org/abs/2012.06356}{{\tt
  arXiv:2012.06356 [gr-qc]}}.

\bibitem{Flathmann:2020zyj}
K.~Flathmann and M.~Hohmann, ``{Post-Newtonian limit of generalized symmetric
  teleparallel gravity},''
  \href{http://dx.doi.org/10.1103/PhysRevD.103.044030}{{\em Phys. Rev. D} {\bf
  103} (2021) no.~4, 044030}, \href{http://arxiv.org/abs/2012.12875}{{\tt
  arXiv:2012.12875 [gr-qc]}}.

\bibitem{Frusciante:2021sio}
N.~Frusciante, ``{Signatures of $f(Q)$-gravity in cosmology},''
  \href{http://dx.doi.org/10.1103/PhysRevD.103.044021}{{\em Phys. Rev. D} {\bf
  103} (2021) no.~4, 044021}, \href{http://arxiv.org/abs/2101.09242}{{\tt
  arXiv:2101.09242 [astro-ph.CO]}}.

\bibitem{Yang:2021fjy}
J.-Z. Yang, S.~Shahidi, T.~Harko, and S.-D. Liang, ``{Geodesic deviation,
  Raychaudhuri equation, Newtonian limit, and tidal forces in Weyl-type
  $f(Q,T)$ gravity},''
  \href{http://dx.doi.org/10.1140/epjc/s10052-021-08910-6}{{\em Eur. Phys. J.
  C} {\bf 81} (2021) no.~2, 111}, \href{http://arxiv.org/abs/2101.09956}{{\tt
  arXiv:2101.09956 [gr-qc]}}.

\bibitem{Fu:2021rgu}
Q.-M. Fu, L.~Zhao, and Q.-Y. Xie, ``{Thick braneworld model in nonmetricity
  formulation of general relativity and its stability},''
  \href{http://arxiv.org/abs/2103.03618}{{\tt arXiv:2103.03618 [gr-qc]}}.

\bibitem{Khyllep:2021pcu}
W.~Khyllep, A.~Paliathanasis, and J.~Dutta, ``{Cosmological solutions and
  growth index of matter perturbations in $f(Q)$ gravity},''
  \href{http://arxiv.org/abs/2103.08372}{{\tt arXiv:2103.08372 [gr-qc]}}.

\bibitem{Anagnostopoulos:2021ydo}
F.~K. Anagnostopoulos, S.~Basilakos, and E.~N. Saridakis, ``{First evidence
  that non-metricity f(Q) gravity could challenge \ensuremath{\Lambda}CDM},''
  \href{http://dx.doi.org/10.1016/j.physletb.2021.136634}{{\em Phys. Lett. B}
  {\bf 822} (2021)  136634}, \href{http://arxiv.org/abs/2104.15123}{{\tt
  arXiv:2104.15123 [gr-qc]}}.

\bibitem{Albuquerque:2022eac}
I.~S. Albuquerque and N.~Frusciante, ``{A designer approach to f(Q) gravity and
  cosmological implications},''
  \href{http://dx.doi.org/10.1016/j.dark.2022.100980}{{\em Phys. Dark Univ.}
  {\bf 35} (2022)  100980}, \href{http://arxiv.org/abs/2202.04637}{{\tt
  arXiv:2202.04637 [astro-ph.CO]}}.

\bibitem{Bernstein:1988ad}
J.~Bernstein, L.~S. Brown, and G.~Feinberg, ``{COSMOLOGICAL HELIUM PRODUCTION
  SIMPLIFIED},'' \href{http://dx.doi.org/10.1103/RevModPhys.61.25}{{\em Rev.
  Mod. Phys.} {\bf 61} (1989)  25}.

\bibitem{Kolb:1990vq}
E.~W. Kolb and M.~S. Turner,
  \href{http://dx.doi.org/10.1201/9780429492860}{{\em {The Early Universe}}},
  vol.~69.
\newblock 1990.

\bibitem{Olive:1999ij}
K.~A. Olive, G.~Steigman, and T.~P. Walker, ``{Primordial nucleosynthesis:
  Theory and observations},''
  \href{http://dx.doi.org/10.1016/S0370-1573(00)00031-4}{{\em Phys. Rept.} {\bf
  333} (2000)  389--407}, \href{http://arxiv.org/abs/astro-ph/9905320}{{\tt
  arXiv:astro-ph/9905320}}.

\bibitem{Cyburt:2015mya}
R.~H. Cyburt, B.~D. Fields, K.~A. Olive, and T.-H. Yeh, ``{Big Bang
  Nucleosynthesis: 2015},''
  \href{http://dx.doi.org/10.1103/RevModPhys.88.015004}{{\em Rev. Mod. Phys.}
  {\bf 88} (2016)  015004}, \href{http://arxiv.org/abs/1505.01076}{{\tt
  arXiv:1505.01076 [astro-ph.CO]}}.

\bibitem{Barrow:2020kug}
J.~D. Barrow, S.~Basilakos, and E.~N. Saridakis, ``{Big Bang Nucleosynthesis
  constraints on Barrow entropy},''
  \href{http://dx.doi.org/10.1016/j.physletb.2021.136134}{{\em Phys. Lett. B}
  {\bf 815} (2021)  136134}, \href{http://arxiv.org/abs/2010.00986}{{\tt
  arXiv:2010.00986 [gr-qc]}}.

\bibitem{Asimakis:2021yct}
P.~Asimakis, S.~Basilakos, N.~E. Mavromatos, and E.~N. Saridakis, ``{Big bang
  nucleosynthesis constraints on higher-order modified gravities},''
  \href{http://dx.doi.org/10.1103/PhysRevD.105.084010}{{\em Phys. Rev. D} {\bf
  105} (2022) no.~8, 084010}, \href{http://arxiv.org/abs/2112.10863}{{\tt
  arXiv:2112.10863 [gr-qc]}}.

\bibitem{Torres:1997sn}
D.~F. Torres, H.~Vucetich, and A.~Plastino, ``{Early universe test of
  nonextensive statistics},''
  \href{http://dx.doi.org/10.1103/PhysRevLett.79.1588}{{\em Phys. Rev. Lett.}
  {\bf 79} (1997)  1588--1590},
  \href{http://arxiv.org/abs/astro-ph/9705068}{{\tt arXiv:astro-ph/9705068}}.
  [Erratum: Phys.Rev.Lett. 80, 3889 (1998)].

\bibitem{Lambiase:2005kb}
G.~Lambiase, ``{Lorentz invariance breakdown and constraints from big-bang
  nucleosynthesis},'' \href{http://dx.doi.org/10.1103/PhysRevD.72.087702}{{\em
  Phys. Rev. D} {\bf 72} (2005)  087702},
  \href{http://arxiv.org/abs/astro-ph/0510386}{{\tt arXiv:astro-ph/0510386}}.

\bibitem{Lambiase:2012fv}
G.~Lambiase, ``{Constraints on massive gravity theory from big bang
  nucleosynthesis},''
  \href{http://dx.doi.org/10.1088/1475-7516/2012/10/028}{{\em JCAP} {\bf 10}
  (2012)  028}, \href{http://arxiv.org/abs/1208.5512}{{\tt arXiv:1208.5512
  [gr-qc]}}.

\bibitem{Lambiase:2011zz}
G.~Lambiase, ``{Dark matter relic abundance and big bang nucleosynthesis in
  Horava's gravity},'' \href{http://dx.doi.org/10.1103/PhysRevD.83.107501}{{\em
  Phys. Rev. D} {\bf 83} (2011)  107501}.

\bibitem{Capozziello:2017bxm}
S.~Capozziello, G.~Lambiase, and E.~N. Saridakis, ``{Constraining f(T)
  teleparallel gravity by Big Bang Nucleosynthesis},''
  \href{http://dx.doi.org/10.1140/epjc/s10052-017-5143-8}{{\em Eur. Phys. J. C}
  {\bf 77} (2017) no.~9, 576}, \href{http://arxiv.org/abs/1702.07952}{{\tt
  arXiv:1702.07952 [astro-ph.CO]}}.

\bibitem{Nakahara:2003nw}
M.~Nakahara, {\em {Geometry, topology and physics}}.
\newblock Taylor and Francis,
2003.
\newblock

\bibitem{Nester:1998mp}
J.~M. Nester and H.-J. Yo, ``{Symmetric teleparallel general relativity},''
  {\em Chin. J. Phys.} {\bf 37} (1999)  113,
\href{http://arxiv.org/abs/gr-qc/9809049}{{\tt arXiv:gr-qc/9809049 [gr-qc]}}.

\bibitem{BeltranJimenez:2019tjy}
J.~B. Jim\'enez, L.~Heisenberg, and T.~S. Koivisto, ``{The Geometrical Trinity
  of Gravity},'' \href{http://dx.doi.org/10.3390/universe5070173}{{\em
  Universe} {\bf 5} (2019) no.~7, 173},
  \href{http://arxiv.org/abs/1903.06830}{{\tt arXiv:1903.06830 [hep-th]}}.

\bibitem{ortin2007gravity}
T.~Ort{\'\i}n, {\em Gravity and Strings}.
\newblock Cambridge Monographs on Mathematical Physics. Cambridge University
  Press, 2007.
\newblock \url{https://books.google.com.mt/books?id=HDmucsxABzYC}.

\bibitem{BeltranJimenez:2018vdo}
J.~Beltr\'{a}n~Jim\'{e}nez, L.~Heisenberg, and T.~S. Koivisto, ``{Teleparallel
  Palatini theories},''
  \href{http://dx.doi.org/10.1088/1475-7516/2018/08/039}{{\em JCAP} {\bf 1808}
  (2018) no.~08, 039},
\href{http://arxiv.org/abs/1803.10185}{{\tt arXiv:1803.10185 [gr-qc]}}.

\bibitem{aldrovandi2012teleparallel}
R.~Aldrovandi and J.~Pereira, {\em Teleparallel Gravity: An Introduction}.
\newblock Fundamental Theories of Physics. Springer Netherlands, 2012.

\bibitem{Harko:2018gxr}
T.~Harko, T.~S. Koivisto, F.~S.~N. Lobo, G.~J. Olmo, and D.~Rubiera-Garcia,
  ``{Coupling matter in modified $Q$ gravity},''
  \href{http://dx.doi.org/10.1103/PhysRevD.98.084043}{{\em Phys. Rev.} {\bf
  D98} (2018) no.~8, 084043},
\href{http://arxiv.org/abs/1806.10437}{{\tt arXiv:1806.10437 [gr-qc]}}.

\bibitem{Bamba:2010wb}
K.~Bamba, C.-Q. Geng, C.-C. Lee, and L.-W. Luo, ``{Equation of state for dark
  energy in $f(T)$ gravity},''
  \href{http://dx.doi.org/10.1088/1475-7516/2011/01/021}{{\em JCAP} {\bf 1101}
  (2011)  021},
\href{http://arxiv.org/abs/1011.0508}{{\tt arXiv:1011.0508 [astro-ph.CO]}}.

\bibitem{Wu:2010av}
P.~Wu and H.~W. Yu, ``{$f(T)$ models with phantom divide line crossing},''
  \href{http://dx.doi.org/10.1140/epjc/s10052-011-1552-2}{{\em Eur. Phys. J. C}
  {\bf 71} (2011)  1552}, \href{http://arxiv.org/abs/1008.3669}{{\tt
  arXiv:1008.3669 [gr-qc]}}.

\bibitem{Ayuso:2021vtj}
I.~Ayuso, R.~Lazkoz, and J.~P. Mimoso, ``{DGP and DGPish cosmologies from
  $f(Q)$ actions},'' \href{http://arxiv.org/abs/2111.05061}{{\tt
  arXiv:2111.05061 [astro-ph.CO]}}.

\bibitem{dvali20004d}
G.~Dvali, G.~Gabadadze, and M.~Porrati, ``4d gravity on a brane in 5d minkowski
  space,'' {\em Physics Letters B} {\bf 485} (2000) no.~1-3, 208--214.

\bibitem{Fairbairn:2005ue}
M.~Fairbairn and A.~Goobar, ``{Supernova limits on brane world cosmology},''
  \href{http://dx.doi.org/10.1016/j.physletb.2006.07.048}{{\em Phys. Lett. B}
  {\bf 642} (2006)  432--435},
  \href{http://arxiv.org/abs/astro-ph/0511029}{{\tt arXiv:astro-ph/0511029}}.

\bibitem{fang2008challenges}
W.~Fang, S.~Wang, W.~Hu, Z.~Haiman, L.~Hui, and M.~May, ``Challenges to the dgp
  model from horizon-scale growth and geometry,'' {\em Physical Review D} {\bf
  78} (2008) no.~10, 103509.

\bibitem{bernstein1989cosmological}
J.~Bernstein, L.~S. Brown, and G.~Feinberg, ``Cosmological helium production
  simplified,'' {\em Reviews of Modern physics} {\bf 61} (1989) no.~1, 25.

\bibitem{Coc:2003ce}
A.~Coc, E.~Vangioni-Flam, P.~Descouvemont, A.~Adahchour, and C.~Angulo,
  ``{Updated Big Bang nucleosynthesis confronted to WMAP observations and to
  the abundance of light elements},''
  \href{http://dx.doi.org/10.1086/380121}{{\em Astrophys. J.} {\bf 600} (2004)
  544--552}, \href{http://arxiv.org/abs/astro-ph/0309480}{{\tt
  arXiv:astro-ph/0309480}}.

\bibitem{Olive:1996zu}
K.~A. Olive, E.~Skillman, and G.~Steigman, ``{The Primordial abundance of He-4:
  An Update},'' \href{http://dx.doi.org/10.1086/304281}{{\em Astrophys. J.}
  {\bf 483} (1997)  788}, \href{http://arxiv.org/abs/astro-ph/9611166}{{\tt
  arXiv:astro-ph/9611166}}.

\bibitem{Izotov:1998mj}
Y.~I. Izotov and T.~X. Thuan, ``{The Primordial abundance of 4-He revisited},''
  \href{http://dx.doi.org/10.1086/305698}{{\em Astrophys. J.} {\bf 500} (1998)
  188}.

\bibitem{Fields:1998gv}
B.~D. Fields and K.~A. Olive, ``{On the evolution of helium in blue compact
  galaxies},'' \href{http://dx.doi.org/10.1086/306248}{{\em Astrophys. J.} {\bf
  506} (1998)  177}, \href{http://arxiv.org/abs/astro-ph/9803297}{{\tt
  arXiv:astro-ph/9803297}}.

\bibitem{Izotov:1999wa}
Y.~I. Izotov, F.~H. Chaffee, C.~B. Foltz, R.~F. Green, N.~G. Guseva, and T.~X.
  Thuan, ``{Helium abundance in the most metal-deficient blue compact galaxies:
  I zw 18 and sbs 0335-052},'' \href{http://dx.doi.org/10.1086/308119}{{\em
  Astrophys. J.} {\bf 527} (1999)  757--777},
  \href{http://arxiv.org/abs/astro-ph/9907228}{{\tt arXiv:astro-ph/9907228}}.

\bibitem{Kirkman:2003uv}
D.~Kirkman, D.~Tytler, N.~Suzuki, J.~M. O'Meara, and D.~Lubin, ``{The
  Cosmological baryon density from the deuterium to hydrogen ratio towards QSO
  absorption systems: D/H towards Q1243+3047},''
  \href{http://dx.doi.org/10.1086/378152}{{\em Astrophys. J. Suppl.} {\bf 149}
  (2003)  1}, \href{http://arxiv.org/abs/astro-ph/0302006}{{\tt
  arXiv:astro-ph/0302006}}.

\bibitem{Izotov:2003xn}
Y.~I. Izotov and T.~X. Thuan, ``{Systematic effects and a new determination of
  the primordial abundance of He-4 and dY/dZ from observations of blue compact
  galaxies},'' \href{http://dx.doi.org/10.1086/380830}{{\em Astrophys. J.} {\bf
  602} (2004)  200--230}, \href{http://arxiv.org/abs/astro-ph/0310421}{{\tt
  arXiv:astro-ph/0310421}}.

\bibitem{gariazzo2021parthenope}
S.~Gariazzo, P.~de~Salas, O.~Pisanti, and R.~Consiglio, ``Parthenope
  revolutions,'' {\em arXiv preprint arXiv:2103.05027} (2021)  .

\bibitem{Arbey:2018zfh}
A.~Arbey, J.~Auffinger, K.~P. Hickerson, and E.~S. Jenssen, ``{AlterBBN v2: A
  public code for calculating Big-Bang nucleosynthesis constraints in
  alternative cosmologies},''
  \href{http://dx.doi.org/10.1016/j.cpc.2019.106982}{{\em Comput. Phys.
  Commun.} {\bf 248} (2020)  106982},
  \href{http://arxiv.org/abs/1806.11095}{{\tt arXiv:1806.11095 [astro-ph.CO]}}.

\end{thebibliography}\endgroup

\end{document}